\documentclass[fleqn,usenatbib]{mnras}

\usepackage{newtxtext,newtxmath}

\usepackage[T1]{fontenc}

\DeclareRobustCommand{\VAN}[3]{#2}
\let\VANthebibliography\thebibliography
\def\thebibliography{\DeclareRobustCommand{\VAN}[3]{##3}\VANthebibliography}

\usepackage{graphicx}	
\usepackage{amsmath}	

\DeclareMathOperator{\sech}{sech}

\title[Stellar bars in co-evolving haloes]{The active role of co-evolving haloes in stellar bar formation}

\author[M. Frosst]{
Matthew Frosst$^{1,2}$\thanks{E-mail: matt.frosst@icrar.org},
Danail Obreschkow$^{1,2}$,
Aaron Ludlow$^{1}$
\\
$^{1}$International Centre for Radio Astronomy Research (ICRAR), University of Western Australia, Crawley, WA 6009, Australia\\
$^{2}$ARC Centre of Excellence for All Sky Astrophysics in 3 Dimensions (ASTRO 3D)\\
}

\date{Accepted XXX. Received YYY; in original form ZZZ}

\pubyear{2023}

\begin{document}
\label{firstpage}
\pagerange{\pageref{firstpage}--\pageref{lastpage}}
\maketitle

\begin{abstract}
We use idealised N-body simulations of equilibrium discs in live and static haloes to study how dark matter co-evolution impacts the assembly of stellar particles into a bar and the halo response. 
Initial conditions correspond to a marginally unstable disc according to commonly used disc stability criteria, and are evolved for the equivalent of about $150$ disc dynamical times ($10$Gyr). 
An extensive convergence study ensures accurate modelling of the bar formation process. 
Live haloes lead to the formation of a strong bar, but the same disc remains unbarred when evolved in a static halo. 
Neither seeded disc instabilities, nor longer ($60$Gyr) simulations result in the formation of a bar when the halo is static. 
When the live halo is replaced with a static analogue at later times the previously robust bar slowly dissipates, suggesting: (1) the co-evolution of the disc and halo is critical for the assembly \textit{and} long-term survival of bars in marginally unstable discs; and (2) global disc stability criteria must be modified for discs in the presence of live haloes. 
In our live halo runs, a ``dark bar'' grows synchronously with the stellar bar. 
Processes that inhibit the transfer of angular momentum between the halo and disc may stabilise a galaxy against bar formation, and can lead to the dissolution of the bar itself. 
This raises further questions about the puzzling stability of observed discs that are marginally unstable, but unbarred.  
\end{abstract}



\section{Introduction}
Bars are found in roughly half of all disc galaxies in the local universe \citep{Eskridge2000,Whyte2002,Marinova2007,Sheth2008,Nair2010,Masters2011}. 
Simulations have explained this prominence by showing that bars quickly assemble in dynamically unstable discs \citep[e.g.][]{Hohl1971, OP1973}. 
Once formed, bars play a key role in galactic evolution by redistributing angular momentum and energy between different radii and galactic components \citep[e.g.][]{TW1984,Athanassoula2003,Dubinski2009,Petersen2019}. 
Furthermore, bars themselves evolve over time, growing in mass, changing in shape, and occasionally buckling \citep{Lokas2019, Kumar2022, Sellwood2020}, ultimately impacting the morphology and kinematics of the galaxy. 

It is now well known that the disc and halo interact to affect the strength, shape, and pattern speed of bars \citep[e.g.][]{Hernquist1992, Debattista2000, Athanassoula2002, Athanassoula2003, HolleyBockelmann2005, Dubinski2009, Collier2018, Petersen2016, Petersen2019}. 
The dynamical state and mass distribution of the halo can also influence bar formation \citep{Saha2013, Long2014, Collier2018, Kumar2022}, leading to an increased interest in using bars to probe the nature and assembly histories of dark matter haloes \citep[hereafter, haloes;][]{Cuomo2020, Roshan2021, Reddish2022, RosasGuevara2022, Kashfi2023, BlandHawthorn2023}. 

However, many early simulations of bar formation modeled the halo or disc/bar component using a static potential. 
For instance, a static potential has been commonly used to mimic the gravitational contribution of the halo when investigating bar formation criteria \citep[e.g.][]{OP1973, ELN1982, Sellwood2001, Sellwood2014} or when decomposing bar orbits \citep[e.g.][]{Binney1982, Contopoulos1989, Athanassoula2002, Saha2012, Molloy2015}. 
Similarly, rigid rotating potentials that mimic bars have been used to study their impact on host haloes \citep[e.g.][]{Weinberg2002, Weinberg2007, HolleyBockelmann2005, Sellwood2008}. 
These models neglect the important transfer of energy and angular momentum between the bar and halo, which co-evolve in a common gravitational field \citep{Athanassoula2003, Dubinski2009, Sellwood2016}. 

Early studies showed that while bars can form rapidly in isolated discs \citep{Hohl1971}, they are suppressed when discs are embedded within static representations of massive sphereical haloes \citep{OP1973, ELN1982, Christodoulou1995}, particularly when the halo is much more massive than the disc. 
This suggests that the formation of a bar only occurs when the disc's local self-gravity dominates that of the halo. 
However, the use of a live (particle-based) halo complicates the picture of bar formation greatly, tending to enable and accelerate the growth rate of bars \citep{Athanassoula2002, Martinez2006}, even in centrally halo dominated discs thought to be bar stable by common bar formation criteria \citep{Sellwood2016, Marasco2018}. 
If bars can form in such stable, sub-maximal discs through interactions with their haloes, then bar formation criteria developed for static haloes must be revised for live haloes \citep{Sellwood2016}, and caution must be applied before using the presence of a bar to infer the properties of a halo. 

\citet{Athanassoula2003} showed that resonant interactions that accelerate bar growth are a two-way process: bar formation is facilitated by an exchange of angular momentum between the disc and halo, and the halo evolves in the presence of the bar. 
The literature remains divided as to whether angular momentum transfer from the bar to the central halo can lead to cored ($\rho \sim r^{0}$) haloes, with some arguing for \citep{Weinberg2002,HolleyBockelmann2005,Weinberg2007,Sellwood2008}, and others against this process \citep{Sellwood2003,McMillan2005,Sellwood2006a,Dubinski2009}. 
Regardless, it is widely agreed that bars drive some evolution of the halo mass and velocity distribution \citep{Tremaine1984, Weinberg1985, Hernquist1992, Athanassoula2007,  Zavala2008, Chan2015, Schaller2015, Petersen2019, Collier2021, Beane2023}, but the impact of the halo evolution on the bar itself remains elusive. 
Changes in the halo distribution function may encourage bar growth, changing the stability of the system \citep{Sellwood2016}. 
Similarly, the importance of the interactions between the disc and halo may vary as the bar itself evolves. 

Complicating matters, simulation choices like the numerical resolution of particle species and gravitational softening length affect galaxy kinematics, disc thickness \citep{BenitezLlambay2018, Ludlow2020, Ludlow2021, Wilkinson2023} and aspects of bar formation \citep[e.g.][]{Weinberg1985, Sellwood2003, Weinberg2007, Dubinski2009}. 
Poorly resolved haloes (less than $10^{6-7}$ particles) can numerically heat disc particles and spuriously increase galaxy sizes and velocity dispersions as particles evolve toward energy equipartition \citep[see][]{Ludlow2021}. 
These effects, present in both cosmological and idealised N-body simulations, have implications for the secular evolution of discs and assembly of bars. 
Quantifying the impact of numerical parameters on bar formation is necessary if simulations are to be used to accurately model bars. 

In this context, we revisit the impact of disc-halo co-evolution on bar formation using idealised simulations of a high-resolution, marginally bar unstable disc galaxy. 
In Sec.~\ref{sec:methods}, we introduce our idealised simulations, and describe our analysis techniques. 
Convergence tests for bar formation covering a range of simulation parameters are presented in Sec.~\ref{sec:resolution}; further discussion about the implications of these tests, particularly with respect to cosmological simulations, are presented in Appendix~\ref{apx:furtherscaling}.  
We contrast bar formation in live and static haloes in Sec.~\ref{sec:results}. 
The transformation of the live halo induced by the galactic bar is discussed in Sec.~\ref{sec:wake}. 
Finally, the results are summarised in Section~\ref{sec:conclusions}. 

\section{Methods}\label{sec:methods}
\begin{figure*}
	\includegraphics[width=\textwidth]{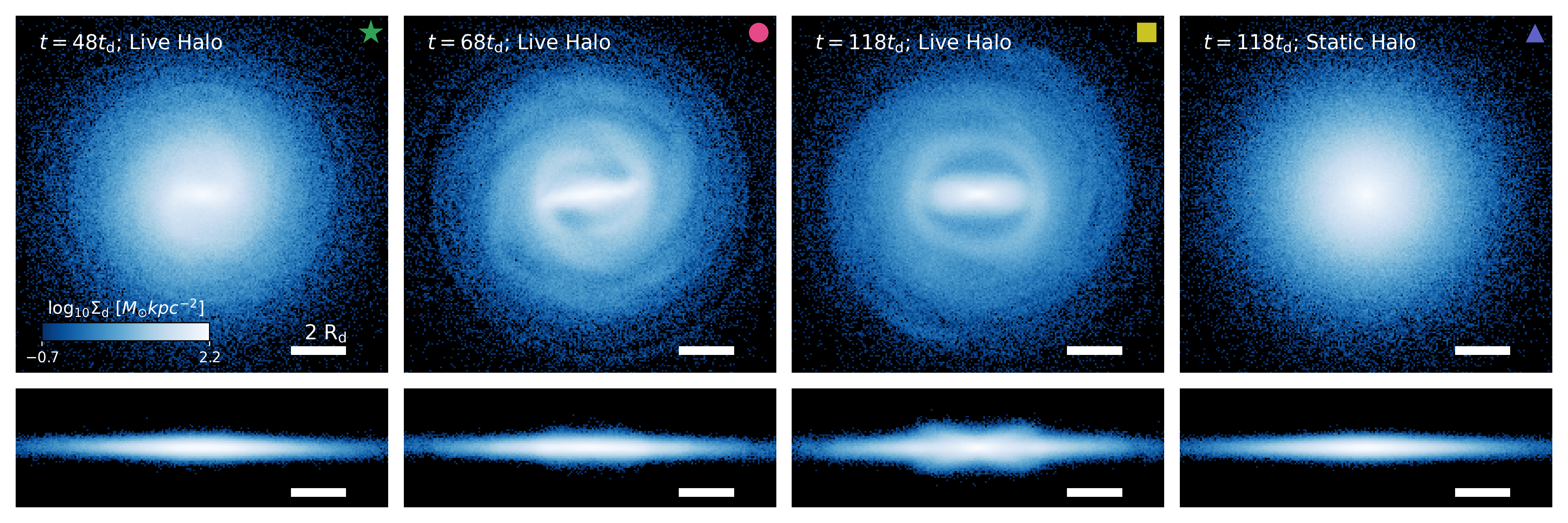}
    \caption{Snapshots of the face-on (top row) and edge-on (bottom row) surface mass density of stars from our fiducial live halo model (first three columns), and identical disc in a static static halo potential (right-most column). 
    Frames are rotated such that the x-axis is parallel to the bar. 
    Each disc has $N_{\rm d} = 10^6$, $N_{\rm h} = 10^7$, $z_{\rm d}/R_{\rm d} = 0.1$. 
    White bars in the bottom right of each panel illustrate the size of $2R_{\rm d}$. 
    Coloured symbols in the upper left of each panel indicate the location of these shapshots in following figures. 
    From left to right, the snapshots progress across the various stages of bar evolution, ``assembly'' at $48t_{\rm d}$ ($\approx3.25$Gyr), ``buckling'' at $68t_{\rm d}$, and ``steady state'' at $118t_{\rm d}$. 
    However, the right-most panels show that the static halo exhibits no evolution over the same timescale.  
    }
    \label{fig:model_example}
\end{figure*}
In this section we present an overview of our how we construct initial conditions, run the simulations, and calculate bar properties. 
Additionally, we perform an in-depth study of the impact of particle mass resolution and gravitational softening on our models. 
In what follows, we adopt a cylindrical coordinate system condiment with the dark matter halo whose $z$-axis is aligned with the disc's angular momentum vector. 
In this coordinate system, $r = (R^2 + z^2)^{1/2}$, where $R=(x^2+y^2)^{1/2}$ is the distance from the $z$-axis and $z$ is the height above the disc. 

\subsection{Initial conditions}\label{sec:ic_generation}
We use \textsc{AGAMA} \citep{Vasiliev2019} to create equilibrium initial conditions (ICs) for our simulations. 
These ICs are created for a disc-halo pair consisting of a \citet{Hernquist1990} halo and thin, rotationally supported stellar disc. 
The disc structure is described by
\begin{equation}
    \rho_{\rm d}(R,z) = \frac{M_{\rm d}}{4 \pi z_{\rm d} R_{\rm d}^2}\exp\left({-\frac{R}{R_{\rm d}}}\right)\sech^2\left({\frac{z}{z_{\rm d}}}\right),
\end{equation}
where $M_{\rm d}$ is the disc mass, and $z_{\rm d}$ and $R_{\rm d}$ are the disc scale height and length respectively. 
The \citet{Hernquist1990} halo structure is described by
\begin{equation}
    \rho_{\rm h}(r) = \frac{M_{\rm h}}{2 \pi} \frac{r_{\rm h}
    }{r(r+r_{\rm h})^3},
\end{equation}
where $r_{\rm h}$ is the halo scale radius, and $M_{\rm h}$ the halo mass.
The \citet{Hernquist1990} profile is similar to a Navarro-Frenk-White profile \citep[NFW;][]{NFW1997} over the radial extent of the disc, but differs at larger radii ($\rho_{\rm h} \propto r^{-4}$ for Hernquist haloes, but $\rho_{\rm h}(r) \propto r^{-3}$ for NFW haloes as $r \rightarrow \infty$). 
For this reason, Hernquist haloes have finite mass, unlike NFW haloes. 
The total mass of our models is $M_{\rm tot} = M_{\rm h} + M_{\rm d}$. 

\textsc{AGAMA} initially assumes that the disc velocity dispersions decline exponentially from the center outwards. 
We set the radial velocity dispersion in the center, $\sigma_{\rm R, 0}$, and fix the velocity dispersion scale radius to $R_{\rm d,\sigma} = 2R_{\rm d}$; these guide the IC generation. 
We use these parameters to set the velocity dispersion profiles and fix the minimum Toomre $Q$ profiles in our discs. 
Our halos have isotropic velocity distributions and no net rotation. 

\textsc{AGAMA} allows users to control the structural properties of the galaxy. 
We specify the disc-to-halo mass fraction $f_{\rm d} = M_{\rm d} / M_{\rm h}$ = 0.03, the ratio of halo-to-disc scale radii $C = r_{\rm h}/R_{\rm d} = 10$, the value of the minimum Toomre Q profile, $\min(Q(R)) = 1.5$, and disc scale height $h_{\rm z} \equiv z_{\rm d}/R_{\rm d} = 0.1$. 
Given the scale-invariance of gravity, systems with identical dimensionless parameters have self-similar solutions irrespective of scale. 
Without loss of generality, we scale our model to be Milky Way-like by setting disc scale length and halo mass to $R_{\rm d} = 2$kpc and $M_{\rm h} = 10^{12} M_{\odot}$, respectively. 
This model sits near the threshold for bar formation in classical stability models \citep{ELN1982}, thus providing both a long bar assembly phase and an interesting comparison of the vigor that live and static haloes lend to bar formation. 

Using the structure of this disc-halo pair, \textsc{AGAMA} computes the combined distribution function (DF) of the system, which we also refer to as the ``fiducial model'', and samples it to produce our ``live halo'' ICs for the N-body simulations.
Alternatively, we sample only the disc DF, and instead replace the live halo with a static \citet{Hernquist1990} potential mimicking the mass distribution of the live halo DF to make a ``static halo'' model. 
By default, we use $N_{\rm d}=10^6$ disc particles, $N_{\rm h} = 10^7$ halo particles (unless the halo is static), and a gravitational softening length, $\epsilon$, of $\epsilon/R_{\rm d}=1/20$ ($\epsilon = h_{\rm z}/2$). 
We establish the convergence and robustness of these values in Sec.~\ref{sec:resolution}. 

Finally, we pin the center of mass to the origin by enforcing axisymmetry in our ICs. 
We do this by duplicating the \textsc{AGAMA} DF with a point-symmetry about the origin, and randomly removing half of all particles. 
This does not affect the DF of the model but has the added benefit of reducing spurious asymmetries in the ICs \citep{Sellwood2024}. 
This requires us to measure the bar strength on only independent disc/halo particles. 

\subsection{Bar analysis}\label{sec:baranalysis}
Bars are quantified by their strength, length, and pattern speed. 
We measure these using a Fourier decomposition of the face-on disc surface mass density distribution \citep[see][]{Athanassoula2002, Guo2019, Dehnen2022}. 
The strength and phase angle of a bar is related to the amplitude of the even Fourier modes, calculated as
\begin{align}
    &A_{\rm m} = |\mathcal{A}_{\rm m}| \text{, and } \\
    &\phi_{\rm m} = \frac{1}{m}\arg (\mathcal{A}_{\rm m}) \text{ where,} \\
    &\mathcal{A_{\rm m}} = \frac{\sum_j M_{j} e^{mi\theta_{\rm j}}}{\sum_j M_{j}},
\end{align}
respectively. 
Here $M_{j}$ and $\theta_{j}$ are the mass and azimuth of the $j$th star particle, and $m$ is the order of the Fourier mode. 
The lowest order radial bar strength profile, $A_{2}(R)$, is measured in cylindrical bins. 
We include exactly $N_{\rm bin} = 10^4$ disc particles per radial bin to keep the Poisson noise, scaling as $\sigma_{A_{2}} = 1 / \sqrt{N_{\rm bin}}$, at $1$ per cent while maintaining good radial resolution. 
While $N_{\rm bin} > 10^4$ better suppresses Poisson noise, for the number of particles in our discs it also leads to poor spatial resolution. 
Tests by \citet{Dehnen2022} indicate that the method with which the galaxy is binned may affect the values of $A_{2}(R)$ and $\phi_{\rm 2}(R)$. 
We present the time evolution of the radial $A_{2}$ profiles of our models in Appendix~\ref{apx:profiles}. 

The bar strength, $A_{2}^{\rm max}$, is defined as the maximum value of $A_{2}(R)$, and has a phase angle defined as $\phi_{2}^{\rm max}$. 
The bar length, $R_{\rm bar}$, is identified as the region around $A_{2}^{\rm max}$ within which $\phi_{2}(R)$ deviates from $\phi_{2}^{\rm max}$ by $\leq \pm 10^{\circ}$ while $A_{2}(R) \geq A_{2}^{\rm max}/2$, following the procedure described in \citet{Dehnen2022}. 
We adopt the definition of \citet{Algorry2017} that systems with $A_{2}^{\rm max} < 0.2$ do not host a bar, while those with $0.2 < A_{2}^{\rm max} < 0.4$ or $A_{2}^{\rm max} > 0.4$ are weakly or strongly barred, respectively. 

Finally, following \citet{RosasGuevara2020}, we define the bar formation time, $t_{\rm bar}$, to be the first time $A_{2}^{\rm max} > 0.2$ and remains so for more than the equivalent of $250$Myr \citep[see also][]{Izquierdo2022}. 
This ensures that we do not erroneously identify bars before they form. 

\subsection{The simulation code}\label{sec:code}
We evolve our models using {\sc GADGET-4} \citep{Springel2021}. 
After testing a variety of gravity solvers, we settled on the Fast Multipole Method (\textsc{FMM}) with a third-order expansion ($p = 3$) because of its superior ability to conserve momentum \citep[originally outlined in][]{Greengard1987}. 
We use the default integration accuracy parameter $\zeta = 0.005$ but have verified that smaller values do not impact the measured bar properties. 

Our disc has a dynamical time of $t_{\rm d} \equiv 2 \pi R_{\rm d}/V_{\rm c}(R_{\rm d})$, where $V_{\rm c}=\sqrt{G M_{\rm tot}(<r) / r}$ is the circular velocity in the ICs, and $M_{\rm tot}(<r)$ is the total enclosed mass. 
In physical units, $t_{\rm d} \approx 68$Myr. 
Given the scale-free nature of gravity, we normalise our time units by $t_{\rm d}$. 
We output snapshots in about $0.74t_{\rm d}$ ($50$Myr) increments for $148t_{\rm d}$ ($10$Gyr), resulting in 200 snapshots per run; higher cadence is not needed to follow the evolution of bar properties. 

\begin{figure}
    \includegraphics[width=\columnwidth]{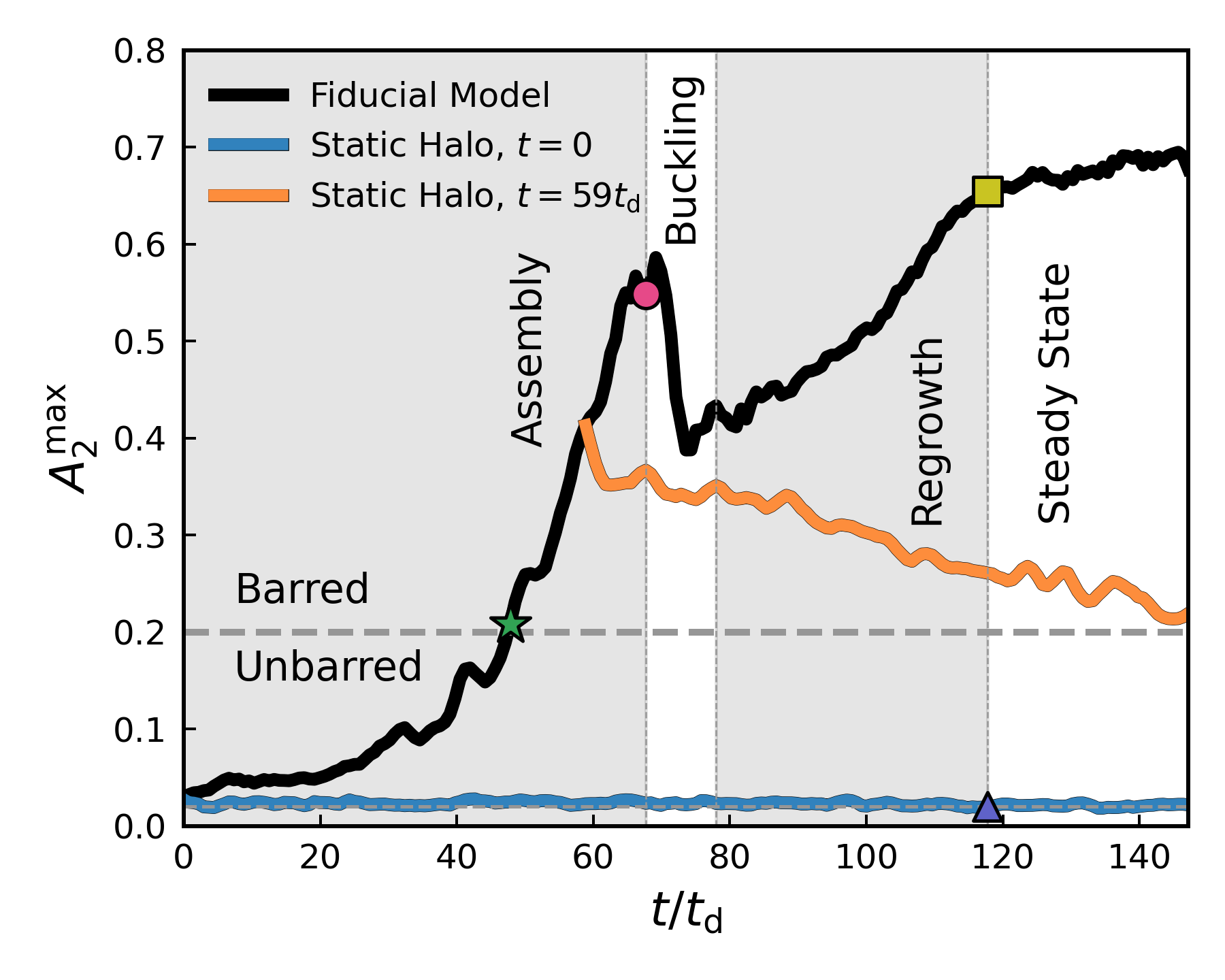}
    \caption{The time evolution of the bar strength of our fiducial model (black line), and the evolution of the disc after the halo halo is frozen in place. 
    We show select static halo runs at $t = [0, 59]t_{\rm d}$ corresponding with the blue and orange lines respectively. 
    The ``assembly'', ``buckling'', ``regrowth'', and ``steady state'' phases of the fiducial model bar evolution are labeled and shown in shaded regions. 
    Coloured points indicate the location of shapshots from Fig.~\ref{fig:model_example}. 
    The thick horizontal dashed grey line shows where $A_{2}^{\rm max} > 0.2$, above which we classify the disc as barred. 
    The lower thinner grey dashed line indicates the Poisson noise floor, $\sigma_{A_{2}}$. 
    }
    \label{fig:freeze_halo}
\end{figure}

In Fig.~\ref{fig:model_example} we show face-on and edge-on disc surface mass density projections for our fiducial live halo (first three panels, from left to right) and static halo model (right-most panel). 
Coloured symbols in the top-right corner of each panel are used to identify these snapshots in later figures. 
We see that the bar in the live halo becomes longer, stronger, and thicker over time, accumulating material in a ring just beyond the bar radius. 
This bar assembles slowly over $\approx68t_{\rm d}$ (about $4.6$Gyr), crossing $A_{2}^{\rm max} > 0.2$ at $\approx48t_{\rm d}$ (first panel). 
The bar then briefly buckles around $\approx68t_{\rm d}$ (second panel from the left), begins to recover from this episode by $\approx80t_{\rm d}$, and finally reaches a steady state at $\approx118t_{\rm d}$ (third panel from the left).  
The same disc does not develop a bar in the static halo model in the same time (rightmost panel; see the full time evolution of the static and live haloes in Fig.~\ref{fig:freeze_halo}). 

\begin{figure*}
    \includegraphics[width=\textwidth]{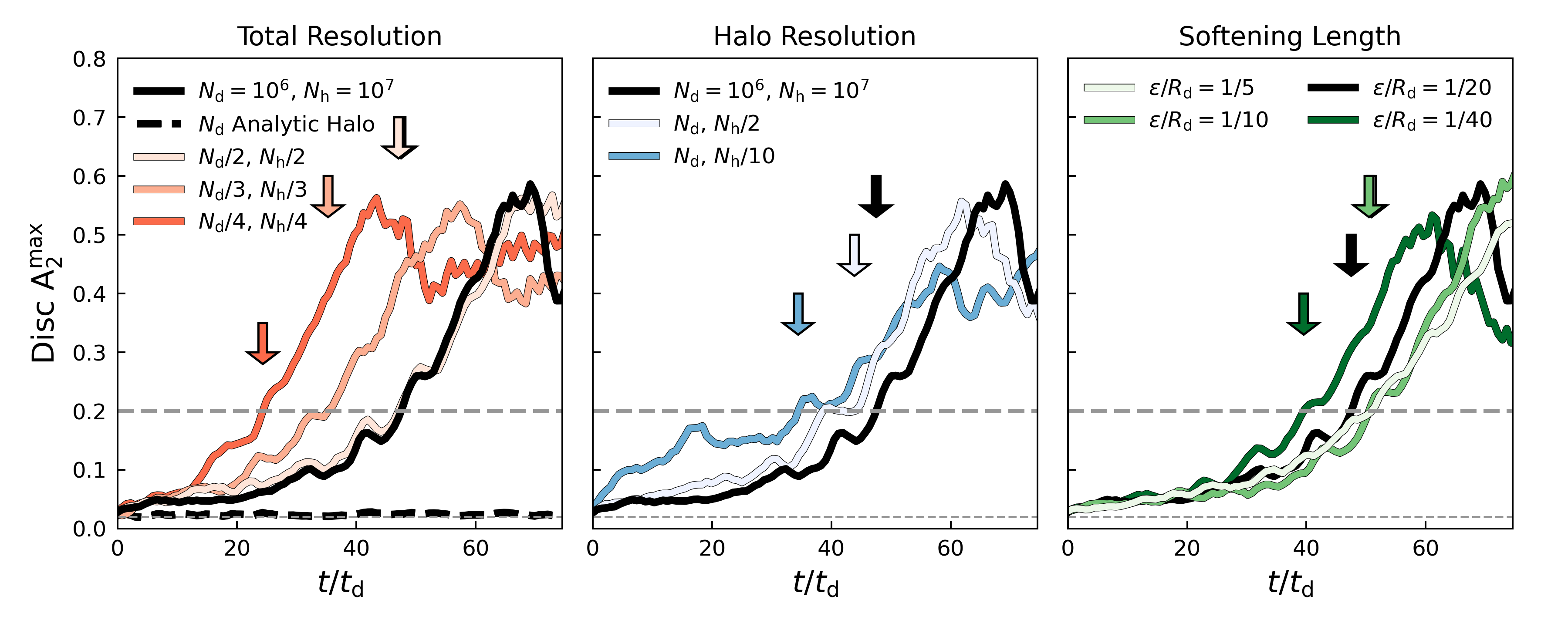}
    \caption{The time evolution of the bar strength given different total particle resolutions (left panel), halo resolutions (middle panel), and softening lengths (right panel). 
    The thick horizontal dashed grey line shows where $A_{2}^{\rm max} > 0.2$, above which we classify the disc as barred. 
    The lower thinner grey dashed line indicates the Poisson noise floor of the fiducial model, $\sigma_{A_{2}}$. 
    In the left panel, the number of total particle resolution is varied as a fraction ($1/2$, $1/3$, and $1/4$; lightest to darkest red respectively) of the fiducial model with $N_{\rm d} = 10^6$ and $N_{\rm h} = 10^7$ (black). 
    The evolution of the fiducial model in a static halo is plotted as a black dashed line. 
    Coloured arrows indicate $t_{\rm bar}$. 
    Simulations converge above half our fiducial resolution; below this threshold, bars form too early if at all. 
    The middle panel shows the impact of halo resolution: each disc has $N_{\rm d} = 10^6$, but $N_{\rm h}$ is varied from fiducial (black) to $N_{\rm h,fid.}/10$ (blue, i.e., halo particles are $\sim30\times$ more massive than disc particles).
    The right panel shows the evolution of the fiducial model with varying the softening-to-scale length ratio; the smallest softening is potted in dark green ($\epsilon/R_{\rm d} = 1/40$), and largest softening in light green ($\epsilon/R_{\rm d} = 1/5$). 
    }
    \label{fig:Nresolution}
\end{figure*}

\section{Numerical Convergence of Bar Formation}\label{sec:resolution}
Before characterising the differences between bar formation in a live versus static halo, we investigate the numerical convergence of our marginally bar unstable model to determine what simulation parameters are required to resolve the earliest stages of bar formation.  
Simulation choices like the number of disc particles, $N_{\rm d}$, the number of halo particles, $N_{\rm h}$, and the gravitational softening length, $\epsilon$, have been shown previously to affect galaxy kinematics, disc thickness \citep{Ludlow2020,Ludlow2021,Wilkinson2023}, and aspects of bar formation \citep{Weinberg1985, Weinberg2007, Dubinski2009, Sellwood2014, Fujii2018}. 
In Fig.~\ref{fig:Nresolution} we present select results of a comprehensive convergence study of these choices. 
We show the evolution of $A_{2}^{\rm max}$ for various re-simulations of the fiducial model with different numerical choices. 
We choose not to show the bar length ($R_{\rm bar}$), pattern speed ($\Omega_{\rm bar}$), or halo $A_{2}$ mode, as these properties show the same convergence behaviour as $A_{2}^{\rm max}$. 

The left-most panel in Fig.~\ref{fig:Nresolution} shows the impact of particle mass resolution on bar formation. 
Here, we show how $A_{2}^{\rm max}$ evolves as we reduce the number of both disc and halo particles relative to fiducial by factors of two, three, and four. 
Halving the number of particles used for the fiducial model ($N_{\rm d}/2=5\times10^5$, $N_{\rm h}/2=5\times10^6$) is sufficient for convergence of the bar assembly phase. 
In general, models with fewer particles develop weaker, longer, more slowly rotating bars after an earlier $t_{\rm bar}$. 
Models with resolution less than $1/50$th the fiducial model resolution will form short lived bars that experience a spurious dampening after buckling. 
We also considered the evolution of an identical disc in a static halo (dashed black line), a distinction which \citet{Sellwood2016} showed increases $t_{\rm bar}$; however, no bar forms in the static halo runs regardless of resolution. 
We discuss the implications of this in Sec.~\ref{sec:results}. 

The central panel of Fig.~\ref{fig:Nresolution} investigates the impact of varying $N_{\rm h}$ at fixed $N_{\rm d}=10^6$. 
In doing so we inadvertently vary the individual halo particle-to-disc particle mass ratio $\mu \equiv m_{\rm h}/m_{\rm d}$, which will have a small impact on disc heating at our mass resolution \citep{Ludlow2021}, and thus bar formation. 
We see that low halo resolution (high $\mu$) will encourage early bar formation even in well resolved discs. 

Finally, the rightmost panel of Fig.~\ref{fig:Nresolution} highlights how changes in the softening length impact the bar evolution for the fiducial mass resolution, with larger (smaller) $\epsilon$ resulting in later (earlier) $t_{\rm bar}$. 
While not plotted, note that there is a complex non-linear relationship between bar formation and softening lengths at different particle mass resolutions \citep[also discussed in][]{Athanassoula2003}; our results are valid only for the fiducial particle mass resolution.   
We settle on an intermediate gravitational softening length of $\epsilon / R_{\rm d} = 1/20$ for our fiducial simulation. 

To ensure the reproducibility of our sample and study the stochastic nature of bar formation, we ran 3 distinct ``seeds'' for the first $59t_{\rm d}$ of the fiducial model, varying the placement and velocities of particles, but keeping the DF fixed. 
The time evolution of disc and bar properties were consistent between different random initializations of the ICs, reassuring us of the robustness of the simulation results against Poisson noise. 

Overall, we find that poorly-resolved discs \textit{or} well-resolved discs in poorly resolved haloes lead to thick and short bar-like structures that rotate much faster than the converged models. 
Furthermore, poor resolution (below $N_{\rm d}/50$ in our tests) can lead to shortened bar lifetimes, as the bars experience spurious dampening. 
The choice of softening length can also change the bar formation time by $\pm 30$ per cent. 
In practice, at least $N_{\rm d}/2=5\times10^5$ disc particles, $N_{\rm h}/2=5\times10^6$ halo particles, and a softening (for this resolution) of $\epsilon/R_{\rm d} \leq 1/20$ is required for convergence \citep[see also][]{Dubinski2009,Sellwood2014, Fujii2018}. 

These requirements may severely bias previous measurements of bars in cosmological simulations which apply fixed softening lengths, and often rely on comparatively poor mass resolution of discs and halos. 
A more detailed account of the impact of disc / halo resolution and softening length on bar formation is found in Appendix~\ref{apx:furtherscaling}. 
Other simulation properties, such as the FMM multipole order, grid size, and opening angle do not appreciably impact the bar, if chosen reasonably. 
Hereafter, our models are run with $N_{\rm d} = 10^6$, $N_{\rm h} = 10^7$, and $\epsilon/R_{\rm d} = 1/20$, above the thresholds required for convergence. 

\section{Bar formation in live and static haloes}\label{sec:results}
We showed in Fig.~\ref{fig:Nresolution} that our fiducial disc model evolving in a live halo forms a bar whereas the same disc in an otherwise identical static halo potential does not. 
Clearly, the disc-halo interaction plays an important role in bar formation. 
We now study how the live halo changes the bar assembly process in this model when compared to the static halo model, and investigate how the bar impacts the evolution of the halo itself. 

In Fig.~\ref{fig:freeze_halo}, we plot the evolution of $A_{2}^{\rm max}$ for our live halo model (black line). 
We identify four distinct stages in the bar evolution: the ``assembly'' phase, where $A_{2}^{\rm max}$ grows exponentially, the ``buckling'' phase beginning at $t=68t_{\rm d}$, where the bar experiences a firehose instability leading to an X-shaped buckled bar (visible in the edge-on projection in the second panel from the left, Fig.~\ref{fig:model_example}), a ``regrowth'' phase, where the bar recovers after the buckling episode, and a ``steady state'' phase at $t \gtrsim 118t_{\rm d}$, where the bar no longer experiences significant changes \citep[similar phases were reported in][]{Petersen2021}. 
The face-on and edge-on projections at several snapshots during each phase are shown in the leftmost panels of Fig.~\ref{fig:model_example} (these projections correspond to locations of the coloured symbols in Fig.~\ref{fig:freeze_halo}). 


To stress the importance of the live halo in bar formation, we replaced the live halo with a static spherical analogue at two distinct times: at the start, $t=0$, and at $t=59t_{\rm d}$, after the bar has formed. 
The static potential deviates from the initial live halo potential by $\leq 0.0025$ per cent at $R_{\rm d}$, and is nearly identical beyond $5R_{\rm d}$; the gradient of the potentials are also very similar. 
We plot the $A_{2}^{\rm max}$ evolution for these static halo models in comparison to the fiducial model in Fig.~\ref{fig:freeze_halo} as blue and orange lines beginning at $t=0$ and $t=59t_{\rm d}$, respectively. 
The resulting bar histories are in stark contrast to the fiducial live halo model: the formation of the bar immediately halts and $A_{2}^{\rm max}$ begins to decline (or never increases). 
The right-most panel of Fig.~\ref{fig:model_example} (dark blue triangle) emphasises that the disc placed in the static halo at $t=0$ has not evolved any a bar even after $t=118t_{\rm d}$. 
We refer to this process as ``bar dissolution'', as it occurs slowly and smoothly over time. 

We reaffirmed these results by carrying out the same tests with a ``frozen'' halo, where halo particles are held fixed during the simulation runtime. 
We find that discs in ``frozen'' halos show a similar bar dissolution as the static spherical halo runs, albeit marginally faster. 
Choosing other times to swap the live halo for a static halo produces similar results. 

The differences in bar histories between our fiducial live halo and static halo models illustrate the importance of the live halo on bar formation. 
These tests suggest that not only is a live halo required to initiate bar formation in marginally bar unstable discs, but it is also necessary for sustaining the bar over long times. 
Note that this is true only in the case of marginally bar (un)stable discs: live and static halos give very similar results when discs are strongly self-gravitating (or non-self gravitating), but only the marginal cases such as those presented here appear to differ substantially.

\subsection{The surprising stability of marginally unstable discs in static haloes} \label{sec:livevsstatic}
\begin{figure}
    \includegraphics[width=\columnwidth]{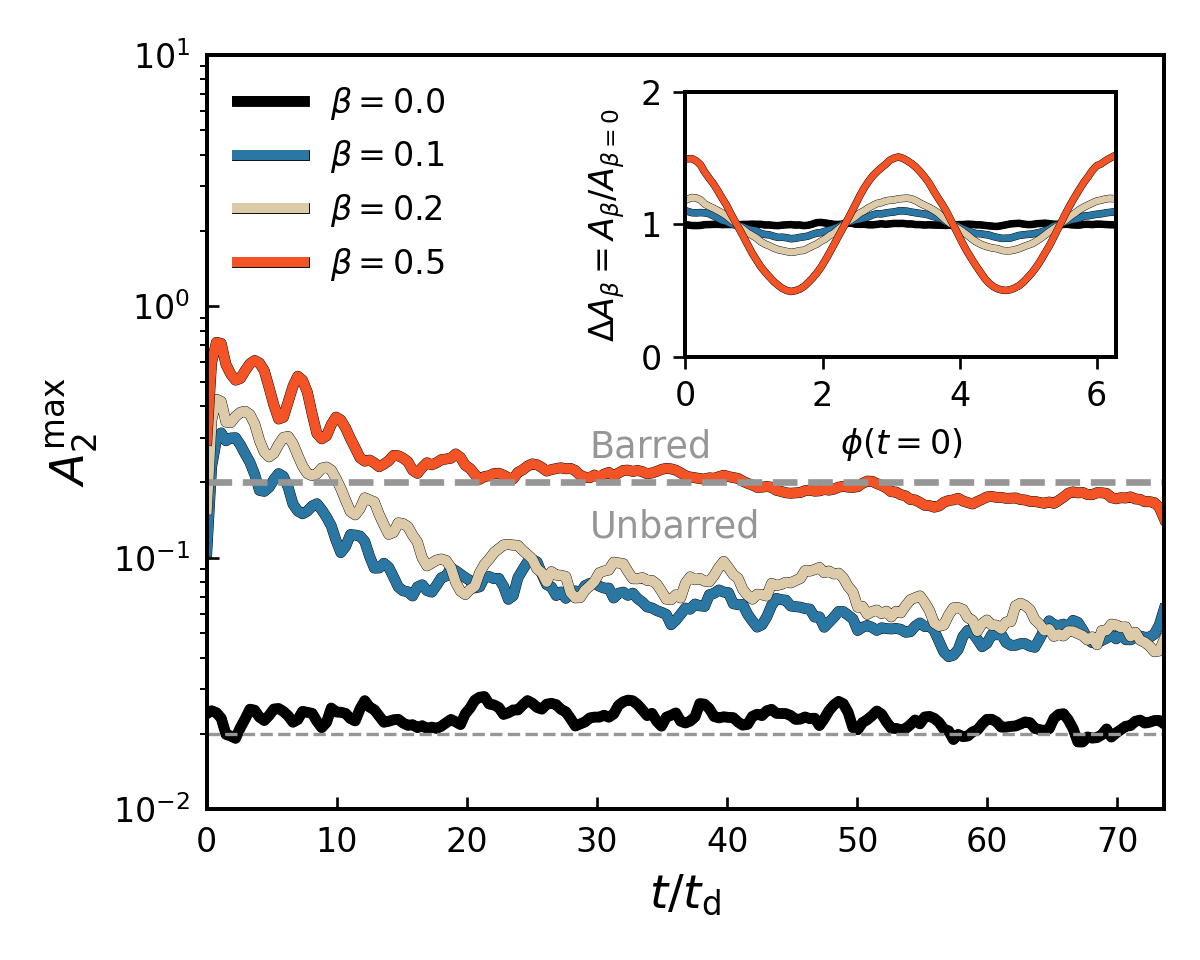}
    \caption{The $A_{2}^{\rm max}$ evolution of a bar unstable model evolved in \textit{static haloes} after being seeded with bar forming instabilities. 
    Models are induced to bar formation by varying the azimuthal distribution of disc particles by a sinusoid of amplitude $\beta$.
    In the inset panel, we show the relative amplitude of particles at a given $\phi$, $A_{\beta}$, normalised by an unperturbed distribution in $\phi$, $A_{\beta=0}$. 
    A model with no induced bar, $\beta = 0$, is plotted as a solid black line, while models with mild, intermediate, and strongly induced bars are plotted in blue, tan, and orange respectively. 
    }
    \label{fig:induced_bar}
\end{figure}

The fact that the bar in our fiducial disc begins to dissipate as soon as the halo is made static suggests that the static halo model is genuinely bar stable. 
To corroborate this, we run two additional tests. 
First, we allow our $t = 0$ model (static halo) to run for $\approx882t_{\rm d}$ ($60$Gyr): no bar forms, and the value of $A_{2}^{\rm max}$ is consistent with Poisson noise.
Second, we induce bar-like perturbations into the disc ICs. 
A ``stable'' disc should develop azimuthal symmetry even if an $m=2$ mode has been seeded. 
We induce $m=2$ perturbations by varying the azimuthal distribution of the disc particles so that they follow a sinusoid of amplitude $\beta$. 
Specifically, for each disc particle with original azimuth $\phi_{\rm i}$, we find a shifted $\phi_{\rm f}$ by solving
\begin{equation}
    \phi_{\rm i} = \phi_{\rm f} + \frac{\beta}{2} \sin(2\phi_{\rm f})
\end{equation}
on the interval $[0, 2\pi]$. 
We vary $\beta$ from $0$ (no induced bar mode) to $0.5$ (which leads to a visible bar-like structure in the ICs and has $A_{2}^{\rm max}$ constant with $R$). 

In Fig.~\ref{fig:induced_bar}, we show the resulting $A_{2}^{\rm max}$ evolution after inducing such bars at $t=0$ in our $t=0$ static halo model. 
We see that $A_{2}^{\rm max}(t=0)$ increases with $\beta$, i.e., the strength of the seeded bar.
Being non-equilibrium ICs, the cases with $\beta > 0$ initially spike up in $A_{2}^{\rm max}$. 
The initial instability is then rapidly damped regardless of the strength of the seeded bar mode. 
In this regard, the fiducial disc is truly stable when evolved in a static halo. 

This differs from the evolution of an identical disc in a live halo which assembles a bar with or without initial bar-mode perturbation. 
The differences are likely driven by two effects which we investigate in Sec.~\ref{sec:wake}. 
First, the live halo distribution contributes small density perturbations to the disc mass. 
And secondly, the halo and bar exchange angular momentum during and after bar formation, further enhancing the growth of bar modes through swing amplification \citep[e.g.][]{Sellwood2014}. 
These mechanisms cannot occur in an axisymmetric potential. 

\subsection{A comment on bar formation criteria}
The differences in bar formation for marginally bar unstable discs in live and static haloes have important implications for bar formation criteria, usually derived for discs in static haloes. 
First efforts to characterize the onset of bar instabilities have shown that dynamically cool, self-gravitating discs are unstable \citep[e.g.][]{OP1973, ELN1982, Christodoulou1995}. 
For example, \citet{ELN1982} developed bar formation criteria from a suite of N-body simulations of razor-thin, exponential stellar discs in static spherical haloes. 
They showed that simulated discs are unstable to bar formation when
\begin{equation}
 \label{eq:eln}
 \epsilon_{ELN} = V_{\rm max} / (GM_{\rm d}/R_{\rm d})^{1/2} \leq 1.1,    
\end{equation}
where $V_{\rm max} \equiv \max\left(\left(G M_{\rm tot}(<r) / r \right)^{1/2} \right)$ is the maximum of the circular velocity of the disc and halo, and $(GM_{\rm d}/R_{\rm d})^{1/2}$ is the velocity of the disc. 
$\epsilon_{\rm ELN}$ is a measure of the importance of self-gravity in the disc: a self-gravitating disc will be more unstable to bar forming modes, and will have a lower $\epsilon_{\rm ELN}$. 
However, the threshold $\epsilon_{\rm ELN} \leq 1.1$ was determined numerically using an insufficient number of particles, and with an unrealistic static halo model \citep[profile in][]{Fall1980}.

Our fiducial model has $\epsilon_{\rm ELN} = 0.97$ and is dynamically cool, so the \citet{ELN1982} stability criteria suggest that this disc should be marginally bar unstable, even in a static halo. 
However, we find it to be stable. 
This may be attributed to: (i) better mass resolution of disc and halo particles (reducing spurious / early bar formation as addressed in Sec.~\ref{sec:resolution} and Appendix~\ref{apx:furtherscaling}), and (ii) using a \citet{Hernquist1990} halo, rather than \citet{Fall1980} halo model \citep[see][]{Syer1998}.

Furthermore, we have shown that the stability of a disc to bar formation depends on whether or not a live halo is used. 
This is not accounted for in previous stability criteria developed for static haloes which under-estimate the stability of our disc in a static halo. 
An array of highly resolved simulations of disc galaxies in live halos is needed to refine past criteria.
We defer updating bar formation criteria to future work, where we present a large scale study of disc secular evolution in a more diverse disc sample. 

\begin{figure}
    \includegraphics[width=\columnwidth]{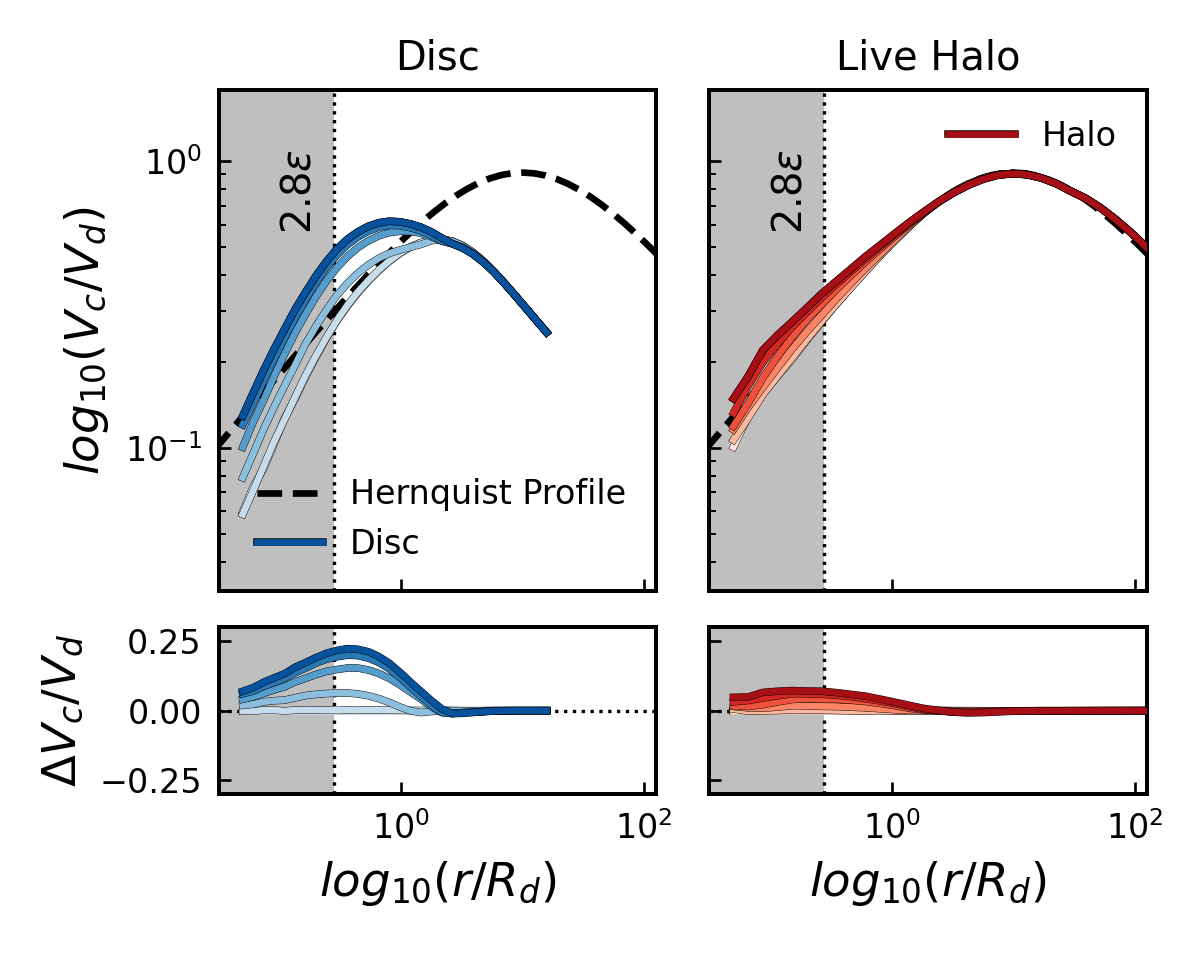}
    \caption{The time evolution of the circular velocity profile (top row) and the residual ($\Delta V_{\rm c} = V_{\rm c}(t) - V_{\rm c}(0)$, bottom row) for a disc situated in a live halo (left column; disc in blue, live halo in red), and the live halo itself (right column). 
    Profiles are plotted at $t \approx [0, 30, 60, 89, 118, 148]t_{\rm d}$; darker colours indicate later times. 
    Profiles are normalised by the initial $V_{\rm d} = \sqrt{G M_{\rm d} / R_{\rm d}}$. 
    The analytic profiles for a isolated Hernquist halo are plotted as thick black dashed lines, while the vertical dotted lines indicate the Plummer equivalent softening length. 
    }
    \label{fig:profile_evolution}
\end{figure}

\begin{figure*}
    \includegraphics[width=\textwidth]{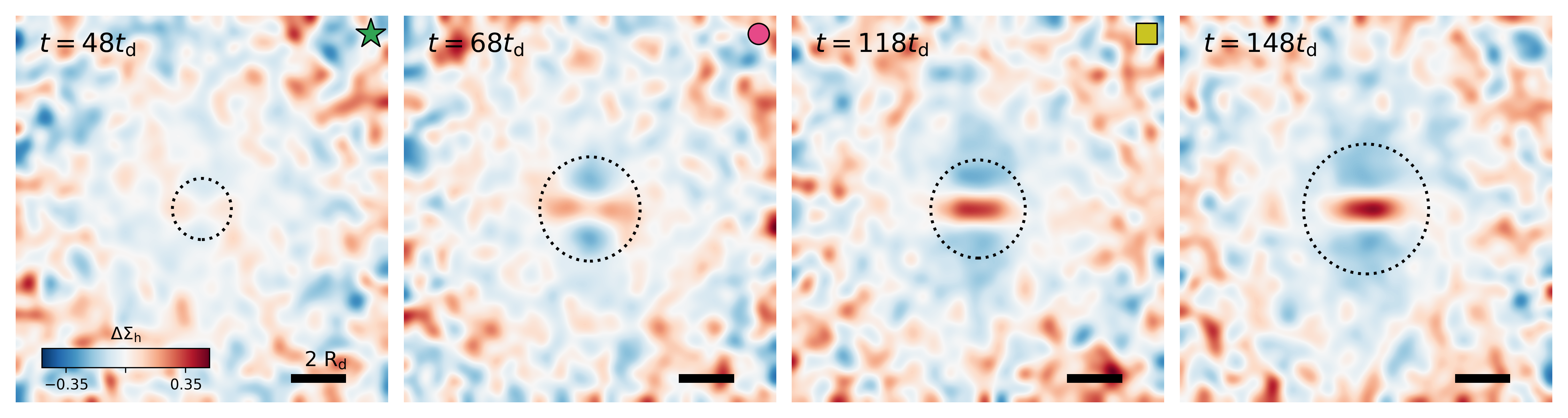}
    \caption{Snapshots of the face-on projection of the relative change in the halo surface density ($\Delta\Sigma_{\rm h} = (\Sigma_{\rm h}(t) - \Sigma_{\rm h}(t=0)) / \Sigma_{\rm h}(t=0)$), highlighting non-asymmetries. 
    Frames are rotated such that the x-axis is parallel to the stellar bar. 
    We plot only a ``slice'' of the halo within $|z| < 2z_{\rm d}$. 
    Dark red indicates a relative over-density of halo of as much as $\times1.5$, while dark blue indicates an under-density of as low as $\times0.5$. 
    Projections are aligned such that the stellar bar is parallel with the x-axis (identically to Fig.~\ref{fig:model_example}). 
    Black bars in the bottom right of each panel illustrate the size of $2R_{\rm d}$. 
    Black dashed circles indicate the length of the associated stellar bar at the same snapshot. 
    Coloured symbols in the upper left of each panel indicate the location of these shapshots in Fig.~\ref{fig:freeze_halo}. 
    From left to right, the snapshots progress across the various stages of bar evolution, the assembly phase at $48t_{\rm d}$, buckling at $68t_{\rm d}$, and steady state at $\approx 118t_{\rm d}$, and a late time snapshot at $\approx148t_{\rm d}$. 
    }
    \label{fig:dm_wake}
\end{figure*}

\begin{figure}
    \includegraphics[width=\columnwidth]{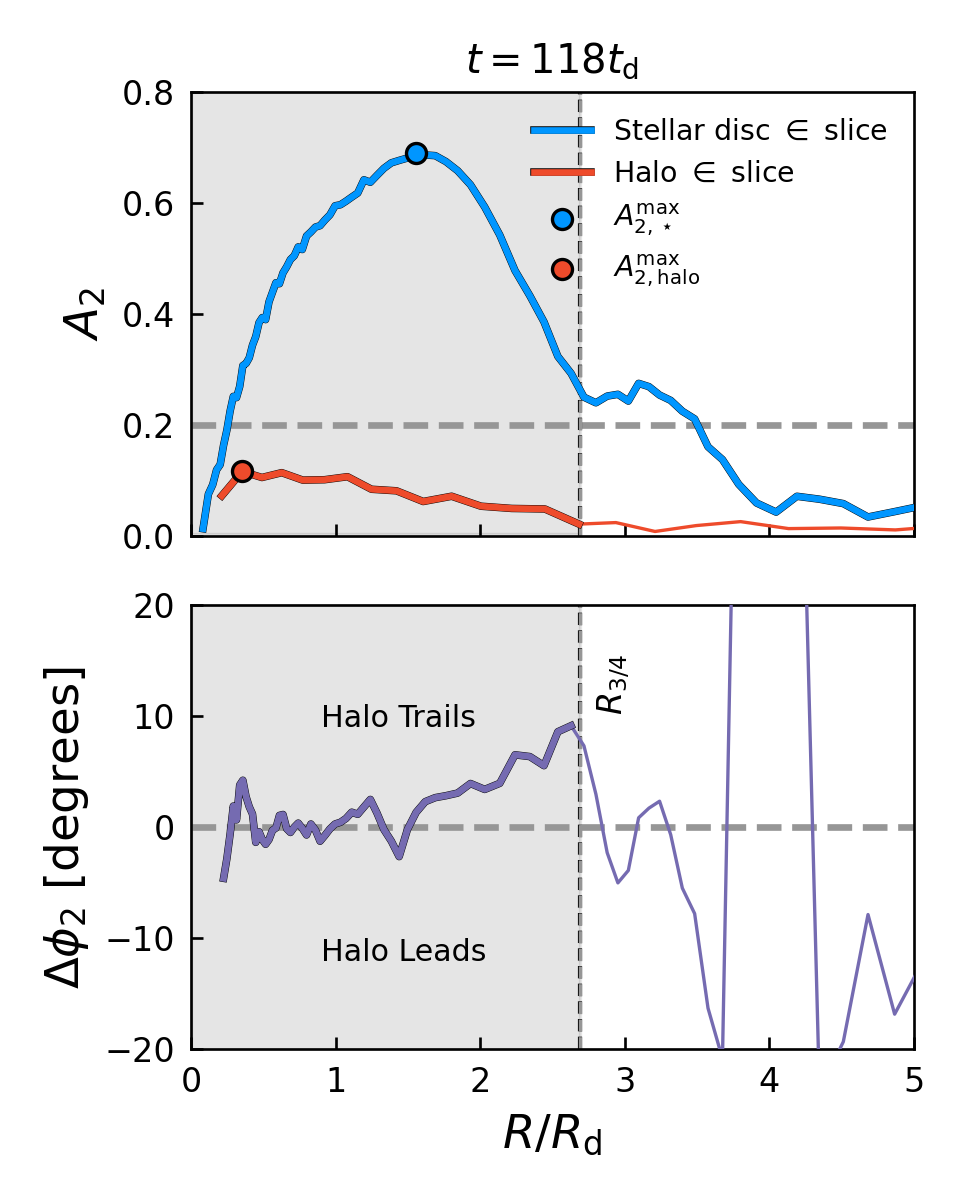}
        \caption{\textit{Upper panel:} The radial $A_{2}$ profile for disc and halo particles within the disc midplane of $|z| < 2z_{\rm d}$ (blue and red lines respectively) at $t=118t_{\rm d}$. 
        The halo $A_{\rm 2, h}^{\rm max}$ and stellar $A_{\rm 2, \star}^{\rm max}$ are shown as red and blue points respectively. 
        \textit{Bottom panel:} The difference in position angle (degrees) between the stellar disc and halo $A_{2}$ as a function of radius. 
        In all panels, the grey shaded region indicates the three quarter-mass radius, $R_{\rm 3/4}$. 
        Lines are thin when Poisson noise dominates. 
        }
    \label{fig:halowake_profile}
\end{figure}

\section{The halo response to bar formation}\label{sec:wake}
We have explored how live and static haloes lead to starkly different bar formation histories in marginally unstable discs (Section~\ref{sec:results}). 
To understand what gives rise to these differences, we next investigate how the live halo evolves in response to the disc evolution. 

\subsection{Halo density evolution}
We show in the upper panels of Fig.~\ref{fig:profile_evolution} that the circular velocity profiles, $V_{\rm c}$, of the disc (blue lines, left panel) and live halo (red lines, right panel) evolve during the simulation. 
As the bar forms, the disc becomes more centrally concentrated and the halo contracts in response. 
Measurements of the mass enclosed within a sphere of radius $R_{\rm d}$ indicate that the central disc mass increases by a factor of $1.64$, and the halo increases by a factor of $1.14$. 
The lower panels of Fig.~\ref{fig:profile_evolution} show the residual change in the circular velocity profile from that of the ICs to emphasise the mass contraction. 
However, when identical halo ICs are allowed to relax in the presence of a static disc potential, the halo mass enclosed within $R_{\rm d}$ contracts by only a factor of $1.08$, indicating that the formation of the stellar bar has enhanced the live halo evolution. 
Ultimately, the halo contraction overcomes any impulse for bar-halo angular momentum exchange to form a core \citep[see also][]{Sellwood2003, McMillan2005, Sellwood2006a, Dubinski2009}. 

\subsection{Identifying the dark bar}
The live halo also develops a triaxial ``dark bar'' near its center. 
\citet{Tremaine1984} predicted this \citep[see also][]{Weinberg1985, Hernquist1992}, and \citet{Athanassoula2007} identified a similar structure in simulations \citep[see also][]{Saha2013, Petersen2016, Petersen2019, Collier2021,Marostica2024}. 
The formation of the dark bar influences the stellar bar, and its structure may also present a challenge in disentangling the influence of baryonic and dark matter in the galactic center \citep{Petersen2016,Marasco2018}. 

We show the fractional change in the face-on halo surface density relative to the ICs in Fig.~\ref{fig:dm_wake} for halo particles enclosed within a midplane volume slice of $z < |2z_{\rm d}|$. 
This volume encloses the stellar bar over the span of the simulation. 
The first three panels show the same outputs used for Fig.~\ref{fig:model_example} and the right-most panel shows the mass distribution at $t=148t_{\rm d}$. 
Projections are aligned such that the stellar bar is parallel to the x-axis. 
The dark bar shows a significant mass over-density parallel to the stellar bar and in spiral-like structures beyond $R_{\rm bar}$ at early times \citep[when $t \lesssim 118t_{\rm d}$, in agreement with][]{Chiba2022} which are the primary source of dynamical friction \citep[see also][]{Saha2013}. 
We also identify relatively less massive under-densities perpendicular to the stellar bar. 
By the end of the simulation, the dark bar has a density contrast of $35$ percent, suggesting a growth rate comparable to that of similar simulations by \citet{Petersen2019}. 
The gravitational potential experienced by these halo particles within the midplane slice also increases by a factor of $\sim1.4\times$ along the stellar/dark bar. 

\begin{figure}
    \includegraphics[width=\columnwidth]{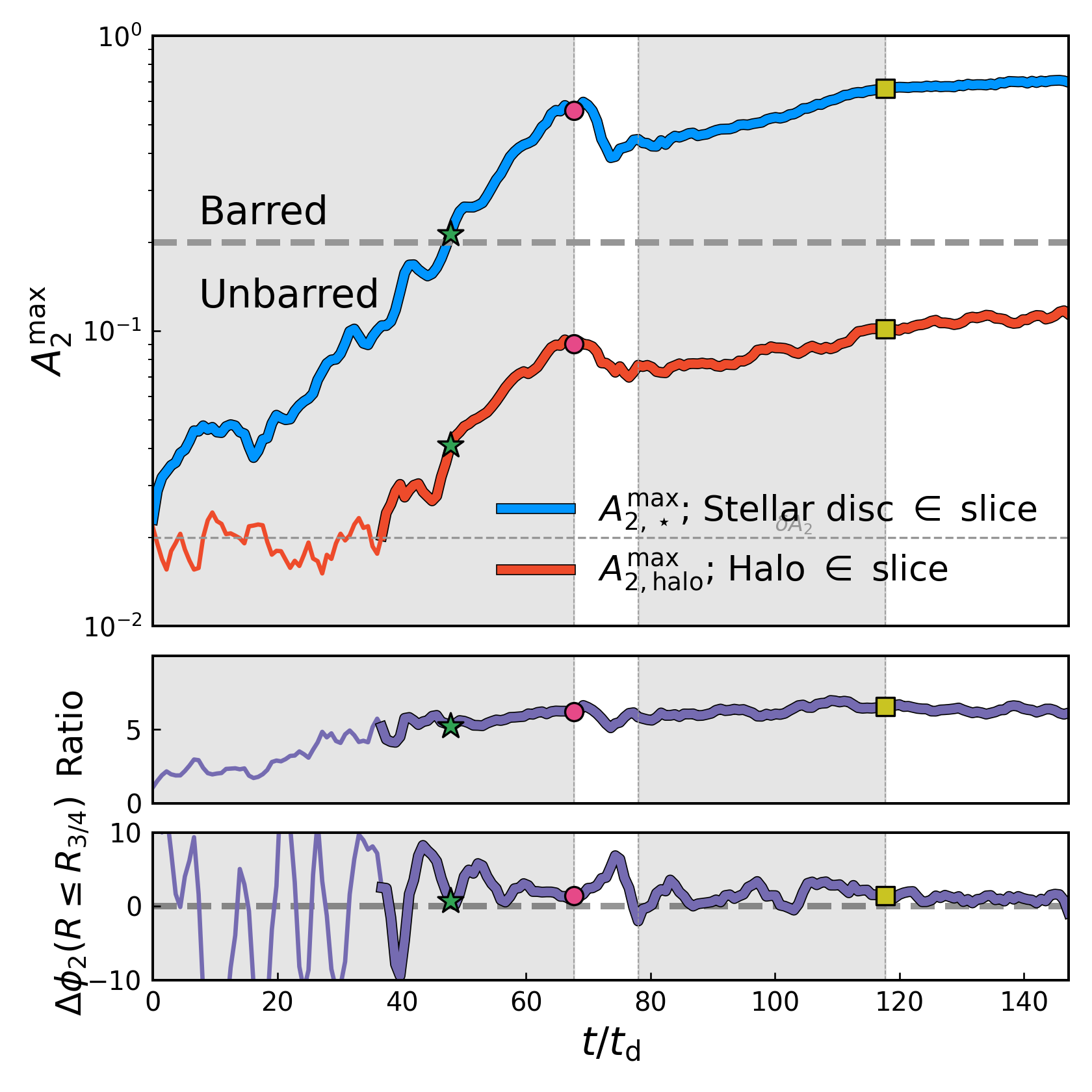}
        \caption{\textit{Upper panel:} The $A_{2}^{\rm max}$ for disc and halo particles within slices of $|z| < 2z_{\rm d}$ and $R < 2R_{\rm d}$ (blue and red respectively). 
        The halo $A_{2}^{\rm max}$ is measured in the radial bin nearest to the radius at which the stellar disc reaches $A_{2}^{max}$. 
        Lines are plotted then when the bar strength is less than the Poisson noise expected for the bin size ($N_{\rm bin}=10^4$ for all particle species). 
        \textit{Middle panel:} The ratio of the disc-to-halo $A_{2}^{\rm max}$ as a function of time, asymptoting to a value of $\sim6$ once both disc and halo measurements are above the noise floor (when lines are no longer thin). 
        \textit{Bottom panel:} The difference in position angle (degrees) between the disc and halo slices as a function of time, measured on all particles within $|z| \leq 2z_{\rm d}$ and $R \leq R_{3/4}$. 
        As in Fig.~\ref{fig:halowake_profile}, lines are thin when near the noise floor. 
        }
    \label{fig:halowake_strength}
\end{figure}

We quantify the strength and position angle of the dark bar using a Fourier decomposition (Sec.~\ref{sec:baranalysis}) on halo particles enclosed within the midplane slice ($z < |2z_{\rm d}|$). 
To improve the comparison with the stellar bar we make similar measurements on stellar particles within the same region. 
The upper panel of Fig.~\ref{fig:halowake_profile} shows the $A_{2}$ profile of the enclosed stellar (blue) and halo (red) particles at $t=118t_{\rm d}$ (after the stellar bar has reached steady state). 
The dark bar strength, $A_{\rm 2, h}$, peaks much closer to the disc center than the stellar bar strength, $A_{\rm 2, \star}$, and is also lower than $A_{\rm 2, \star}$ at all radii. 
The shaded region indicates the radii below the stellar three-quarter mass radius, $R_{3/4}$. 
Within $R_{3/4}$, $A_{\rm 2, h}$ is considerably higher than the expected Poisson noise floor and therefore is reliably obtained from a Fourier decomposition. 

While the strength of the stellar and dark bars are important, it is the degree of alignment between the two that dictates their ability to torque one another and exchange angular momentum \citep{Athanassoula2003}. 
In the lower panel of Fig.~\ref{fig:halowake_profile}, we show the difference between the position angles of the stellar and dark bars, $\Delta \phi_{2} = \phi_{\rm 2, h} - \phi_{\rm 2, \star}$, as a function of radius. 
At low radii, the stellar and dark bars are aligned ($\Delta \phi_{2} \approx 0$), but at radii greater than the peak of $A_{\rm 2, \star}^{\rm max}$, the halo begins to trail behind the stellar bar \citep[see also][]{Athanassoula2007, Saha2013, Petersen2016}. 
The lag increases linearly with $R$ to a maximum deviation of $\Delta \phi_{2} \approx 10$ degrees at $R_{\rm 3/4}$, beyond which the properties of the dark bar become Poisson noise dominated. 
Thickening the midplane slice leads to marginally higher $\Delta \phi_{2}$ at higher $R$: the dark bar leaves a wake that trails the stellar bar not only as a function of $R$, but also in $z$. 

\begin{figure}
    \includegraphics[width=\columnwidth] {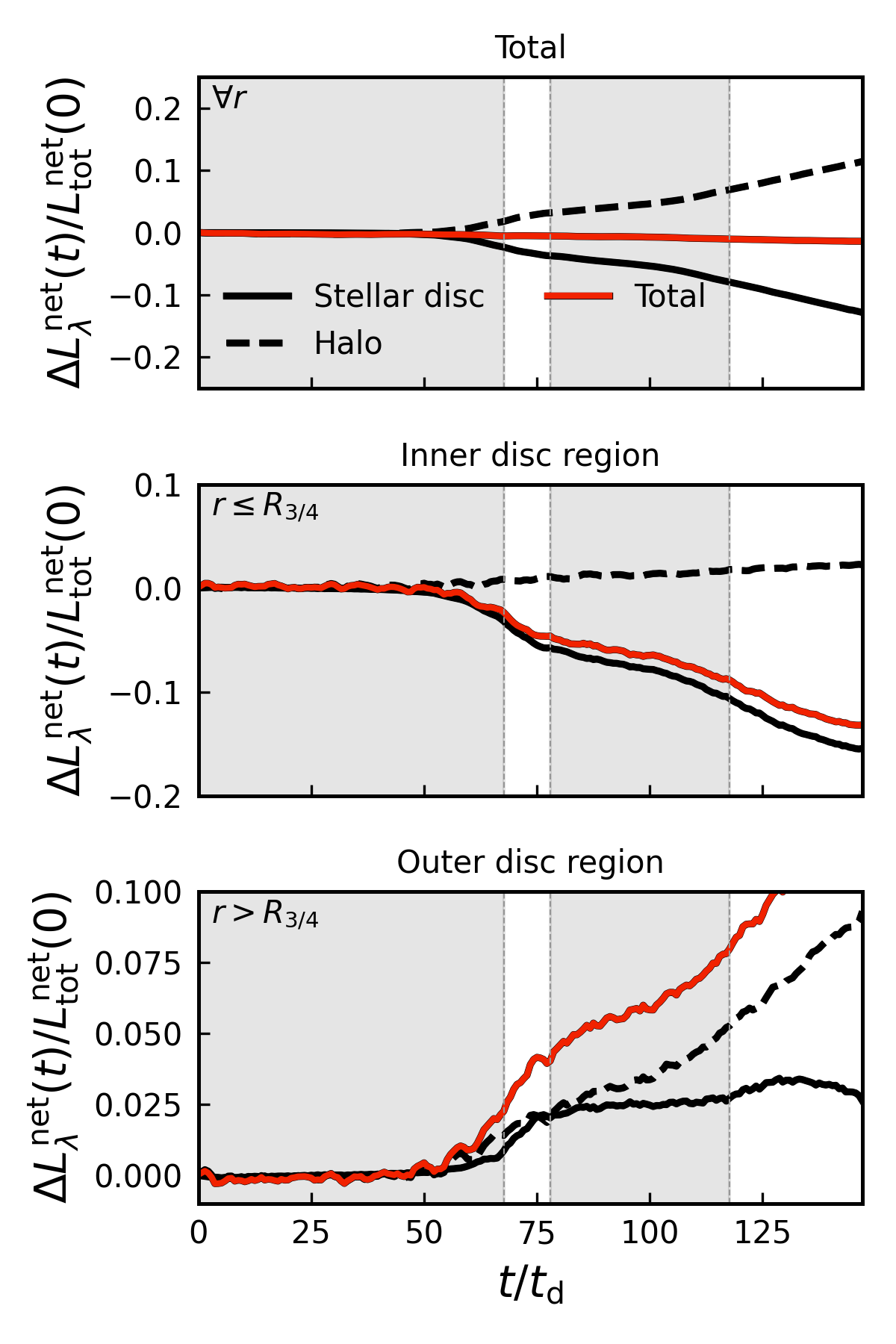}
    \caption{The time evolution of the net (normalised) angular momentum, $\Delta L_{\lambda}^{\rm net}(t)$, for the total system (red line), disc particles (solid black line), and halo particles (dashed black line). 
    From top to bottom, the panels show $\Delta L_{\lambda}^{\rm net}(t)$ for all particles, particles within $r \leq R_{3/4}$, and particles beyond $R_{3/4}$. 
    Shaded regions denote the bar assembly, buckling, regrowth, and steady state phases from left to right respectively. 
    }
    \label{fig:AM_evolution}
\end{figure}

\subsection{The coupling of the dark and stellar bars}
In Fig.~\ref{fig:halowake_strength}, we show the stellar and dark bar properties as a function of time, measured on all particles within $|z| < 2z_{\rm d}$ and $R \leq R_{3/4}$. 
In the upper panel, we show that $A_{2, \star}^{\rm max}$ and $A_{\rm 2, h}^{\rm max}$ evolve simultaneously, even when no strong stellar bar is present (i.e., when $A_{\rm 2, \star}^{\rm max} < 0.2$). 
The times at which the stellar bar is strongest correspond to the strongest dark bars. 
Quantitatively, the exponential bar formation timescale, $\tau_{\rm bar}$ (estimated by fitting $A_{\rm 2, i}^{\rm max}(t)$ to an exponential of the form $ \propto \exp(t/\tau_{\rm bar})$), is very similar for the disc and dark bars with $\tau_{\rm bar, \star} = 1.25t_{\rm d}$ and $\tau_{\rm bar, h} = 1.4t_{\rm d}$, respectively. 

The stellar-to-dark bar strength ratio $A_{2, \star}^{\rm max} / A_{\rm 2, h}^{\rm max}$ is plotted in the middle panel of Fig.~\ref{fig:halowake_strength}. 
At $t \lesssim 48t_{\rm d}$, $A_{\rm 2, h}^{\rm max}$ is Poisson noise dominated, and quantifying the relative strength of the disc and dark bars is difficult. 
However after the bar assembly phase, when the strengths of $A_{\rm 2, \star}^{\rm max}$ and $A_{\rm 2, h}^{\rm max}$ have sufficiently increased, the ratio $A_{\rm 2, \star}^{\rm max} / A_{\rm 2, h}^{\rm max}$ reaches $\approx6$ and remains constant. 
This is true even during the buckling phase of the stellar bar, as the dark bar shows a commensurate decline in $A_{\rm 2, h}^{\rm max}$ in response to the weakening of the stellar bar. 
While the asymptote at $A_{\rm 2, \star}^{\rm max} / A_{\rm 2, h}^{\rm max} \approx6$ is unique to this model, this is further evidence that the stellar and dark bars evolve simultaneously, and are best thought of as a single structure. 

The bottom panel of Fig.~\ref{fig:halowake_strength} shows the evolution of the difference in position angle, $\Delta \phi_{2}$, between the stellar and dark bars. 
After bar assembly when the stellar and dark bars are no longer Poisson noise limited, the dark bar trails the stellar bar position angle by $\Delta \phi_{2} \approx 5$ degrees, oscillating to values as high as $\approx 10$ degrees only during bar buckling. 
Measurements at higher radii show larger $\Delta \phi_{2}$, but are more noise dominated. 
The position angle offset decreases as a function of time to $\Delta \phi_{2} \approx 1$ at $t=148t_{\rm d}$. 
The stellar and dark bars cannot exert torques on each other when they are aligned, however even a small offset slows down the bar pattern speed over time. 
Ultimately, the strong alignment we find between the dark bar and stellar bar at late times implies that there may be less torque than expected by linear theory alone \citep[][]{Petersen2016, Chiba2022, Hamilton2023}.  
Others have reported more significant deviations between the stellar and dark bar components \citep[e.g.,][]{Beane2023}, but even some cosmological simulations find $\Delta \phi_{2} \leq 10$ in marginally bar unstable systems \citep{Marasco2018}. 

In Fig.~\ref{fig:AM_evolution} we plot the net change in angular momentum of the disc and halo, $\Delta L_{\lambda}^{\rm net} = L_{\lambda}^{\rm net}(t) - L_{\rm tot}^{\rm net}(0)$, where the subscript $\lambda$ denotes either a measurement on disc or halo particles, normalised by the total initial net angular momentum of the system, $L_{\rm tot}^{\rm net}(0)$. 
\citet{Weinberg1985} showed that the halo acts as an important angular momentum sink during bar formation \citep[see also][]{Athanassoula2003}. 
We bin the system into two radial shells: an inner disc ($R \leq R_{3/4}$), and outer disc region (otherwise). 

First, the upper panel of Fig.~\ref{fig:AM_evolution} shows that the total angular momentum is conserved throughout the simulation (i.e., the red line is constant) and that the disc loses angular momentum to the halo as expected \citep[e.g., ][]{LyndenBell1972, Athanassoula2003}. 
Second, we see that the degree of angular momentum exchange follows the bar formation phases identified with $A_{2}^{\rm max}$ in Sec.~\ref{sec:results}: the net exchange increases during the assembly phase, levels off briefly during the buckling phase, and increases again near the steady state phase. 
This link between the rate of angular momentum exchange and the stage of the bar formation history appears regardless of radial bin. 
In the inner disc region, the amount of angular momentum exchange levels off as the stellar and dark bar become more closely aligned (see Fig.~\ref{fig:halowake_strength}). 
Finally, while the halo gains angular momentum at all radii, the disc only loses it in the inner disc region, where the stellar bar is forming. 
Similarly, the halo gains far less angular momentum in the inner disc region than in the outer disc region, as some angular momentum is lost to form the dark bar. 
A primary difference between our live and static halo models is that supplanting the live halo with a static potential abruptly halts the transfer of angular momentum between the two components. 

\section{Summary and Conclusions}\label{sec:conclusions}
We used idealised N-body simulations of a marginally unstable disc galaxy to investigate the role played by a live dark matter halo in the formation and long-term survival of a stellar bar. 
We showed that a bar can form when the disc evolves in a live halo, but that it fails to form a bar in a static halo with an identical mass distribution. 
Furthermore, when the live halo is replaced by a static analogue the bar slowly dissipates, and the disc regains axi-symmetry, regardless of its previous evolutionary stage. 
The key result of our study is that facilitating bar assembly is not the only role of the live halo: the live halo also stabilises the bar once formed. 
Our main results are summarised as follows:
\begin{enumerate}
    \item the simultaneous evolution of the disc and halo is critical for the formation \textit{and} sustainment of bars in unstable discs;
    \item  stellar bars form more readily in discs suspended in live haloes than static ones; and,
    \item The stellar bar forms simultaneously with an associated ``dark bar'' in the galactic center, which grows and decays at the same rate as the stellar bar. 
    \item The torques between the stellar and dark bar lead to angular momentum exchange between the disc and halo, a mechanism not present when the halo is modeled as an static potential. 
\end{enumerate}

We have also presented convergence tests in Sec.~\ref{sec:resolution} and Appendix~\ref{apx:furtherscaling}, which show that our simulations are reproducible and sufficiently resolved to converged bar properties. 
Furthermore, the effects of numerical heating are negligible at our fiducial resolution. 
A minimum of $5\times10^5$ disc particles are needed to model the bar assembly phase accurately, in agreement with the earlier estimates of \citet{Dubinski2009}. 
Similarly, we find at least $5\times10^6$ halo particles are required to converge bar properties. 
Lower mass resolution spuriously accelerate bar formation, shorten bar lifetimes, and lead to thick, short bar-like structures that rotate faster than the converged solution. 
Poor choices of gravitational softening length can vary the bar formation time by $\pm 30$\%. 
We have invested significant effort in ensuring our ICs are in equilibrium, testing three separate codes, and settling on the \textsc{AGAMA} software. 
No indication of ``disc settling'' occurs when running our simulations. 

Our simulations neglect star formation, energetic feedback, galactic components such as bulges and stellar halos, substructure, gas discs, and other elements necessary align simulations with observations. 
This simplicity allow us to better understand the dynamical processes relevant for bar formation, but limits their applicability to real galaxies. 
In a companion paper, we will address some of these issues by investigating secular bar formation across a more diverse range of disc galaxy models. 

\section*{Acknowledgements}
MF acknowledge support from the Australian Government
Research Training Program (RTP) Scholarship. 
ADL and DO acknowledge financial support from the Australian Research Council through their Future Fellowship scheme (project numbers FT160100250, FT190100083, respectively). 
This research was undertaken with the assistance of resources and services from the National Computational Infrastructure (NCI), which is supported by the Australian Government.
This work has benefited from the following public \textsc{Python}
packages: \textsc{AGAMA} \citep{Vasiliev2019} \textsc{Scipy} \citep{Virtanen2020}, \textsc{Numpy} \citep{Harris2020}, and \textsc{Matplotlib} \citep{Hunter2007}.

\section*{Data Availability}

Our data is available upon reasonable request, or is otherwise obtainable from publicly available codes. 
 


\bibliographystyle{mnras}
\bibliography{references} 


\appendix
\section{Further Numerical Convergence Tests}\label{apx:furtherscaling}

\subsection{The impact of halo resolution}\label{apx:mu} 
The work of \citet{Ludlow2021} and later \citet{Wilkinson2023} have shown that halo resolution can impact the thickness of galaxies through collisional interactions between disc and halo particles. 
These collisions result in spurious thickening of the stellar disc, as the system moves towards energy equipartition by exchanging energy and angular momentum between the stellar and halo components. 
This effect is more pronounced at lower $N_{\rm h}$, but always present as a phase-space segregation between components. 
The choice of $N_{\rm h}$ thus impacts galaxy morphology and kinematics, and it is expected that it will also impact bar formation, as \citet{Klypin2009} has shown that thicker discs are more stable to global instabilities \citep[see also][]{Aumer2017, Ghosh2022}. 
We further justify our choice of $N_{\rm h} = 10^7$ by running 12 additional simulations with fixed $N_{\rm d} = 10^6$ and varied $N_{\rm h}$ probing $\mu \equiv m_{\rm h}/m_{\rm d} = 1, 5, 10, 20, 30, 40, 50, 75, 100, 200, 250, 500$. 
Given that $f_{\rm d}$ is fixed, this is equivalent to varying $N_{\rm h}$ from $3\times10^7$ to $6\times10^4$. 

We find that if a bar forms in the disc, the initial stage of the bar evolution is exponential regardless of the value of $N_{\rm h}$ (see middle panel, Fig.~\ref{fig:Nresolution}). 
However, the bar strength growth rate is decreased by a factor of 2 when $N_{\rm h}$ decreases by 2 orders of magnitude. 
The peak of the bar strength is also marginally lower at lower $N_{\rm h}$. 
Of note, the degree of stochasticity in bar formation (the variation between different random samples from the same DF) does not seem to be affected by varying halo resolution. 

We also find a coherent trend between $N_{\rm h}$ and other bar properties, particularly when $N_{\rm h} \leq 10^6$ ($\mu \geq 10$). 
If $N_{\rm h}$ is chosen below this threshold, bars form too long and rotate too slowly. 
Ultimately, all bar properties are converged when $N_{\rm h} \geq 5\times10^6$ ($\mu \lesssim 5$). 
Therefore, if chosen reasonably, $N_{\rm h}$ has little impact on the formation of the bar; we shall see that the disc particle resolution is far more important. 
This is expected in light of \citet{Ludlow2021}, who find that the increase in the velocity dispersions and scale height of stellar particles depends only weakly on $\mu$. 
Ultimately, both the rate of collisional heating and the growth of bar modes are driven by stochastic fluctuations in the galaxy potential which lead to orbital deflections in the stellar particles that also depend only weakly on $\mu$. 

\subsection{The impact of gravitational softening length}\label{apx:softeninglength}
The simulations presented in this study were run with a variable softening length of $\epsilon/R_{\rm d} = 1/20$. 
However, \citet{Ludlow2020, Ludlow2021} highlight the important effects that the softening length $\epsilon$ have on galaxy morphology. 
They found that if the softening length is too large small-scale clustering of stellar and DM particles will be reduced, resulting in galaxies that are underdense and too large, with artificially induced DM cores. 
On the other hand, if the softening length is too small, the impact of collisional heating on the stellar disc will be intensified, leading to spurious evolution of the disc properties (such as numerically induced increases in scale heights, velocity dispersions, etc.). 
Bar formation introduces a characteristic length scale that must be resolved or the onset of instabilities will fail to destabilise the disc. 
It is therefore expected that the choice of softening length will play an important role in the simulated evolution of barred galaxies, such as those investigated in this study.

To determine the impact of the softening length on bar formation, we run four identical models ($N_{\rm d} = 10^6$ and $N_{\rm h} = 10^7$, the fiducial resolution) with softening lengths varying from $\epsilon/R_{\rm d} = 1/40$ to $\epsilon/R_{\rm d} = 1/5$. 
We show the time evolution of the bar properties of these runs in the rightmost panel of Fig.~\ref{fig:Nresolution}. 
Runs with the lowest softening length are shown as the darkest green lines, while runs with the largest softening are shown as the lightest green lines. 
Coloured arrows indicate the bar formation time. 
This figure makes clear our choice of $\epsilon/R_{\rm d} = 1/20$. 
Larger softening lengths lead to delayed bar formation, particularly apparent when $\epsilon/R_{\rm d} > 1/10$. 
\citet{Ludlow2021} also find that velocity dispersion and scale height profiles of simulated disc galaxies are independent of $\epsilon$ above this threshold. 

\begin{figure}
    \includegraphics[width=\columnwidth]{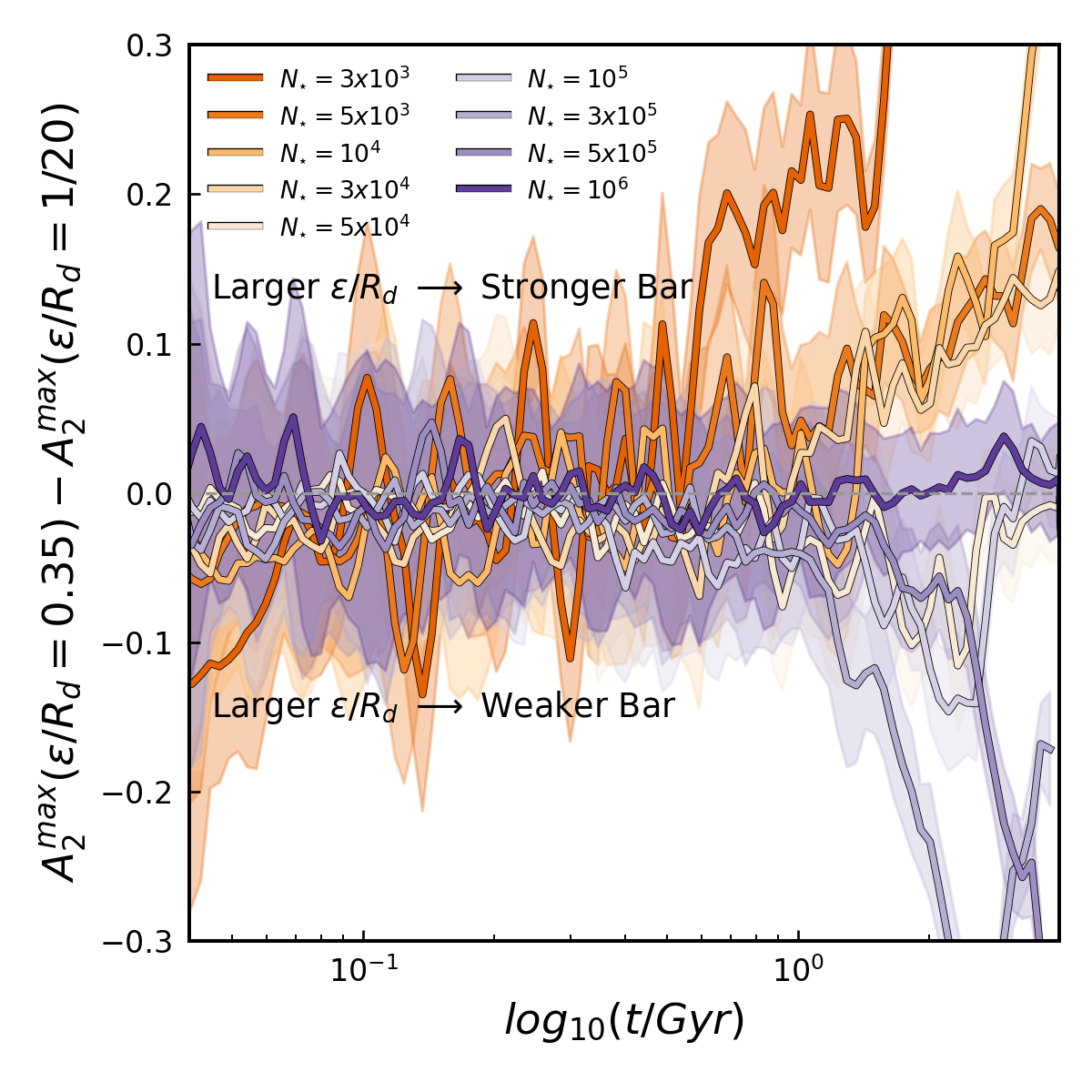}
    \caption{The difference in bar strength between runs with $\epsilon/R_{\rm d} = 0.35$ and $\epsilon/R_{\rm d} = 1/20$ over $4$Gyr.  
    Residuals are plotted for different model resolutions, all with $\mu=5$, from $N_{\rm d}=3\times10^3$ (dark orange) to $N_{d}=10^6$ (dark purple). 
    At resolutions below $N_{\rm d} = 3\times10^4$, large softening leads to stronger bars, while at intermediate resolutions $5\times10^4 < N_{\rm d} < 10^6$, large softening delays bar formation. 
    When resolution is sufficiently high $N_{\rm d} \geq 10^6$, the differences between different $\epsilon$ have minimal impact on the bar formation process. } 
    \label{fig:softening_scaling}
\end{figure}

\begin{figure*}
    \includegraphics[width=\textwidth]{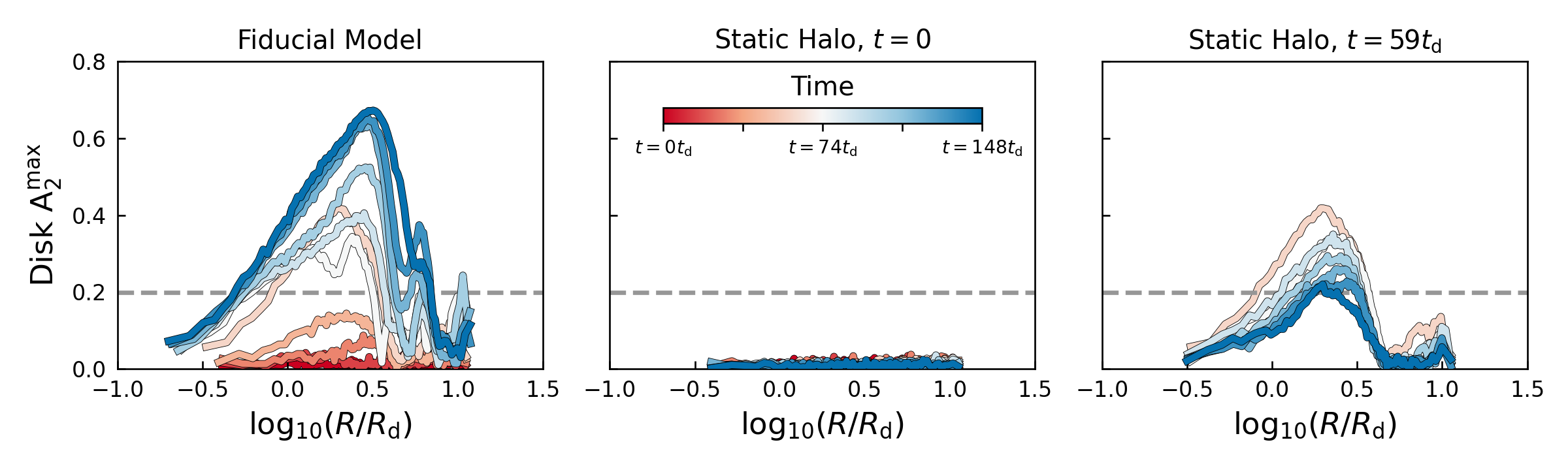}
    \caption{The evolution of the radial $A_{2}$ profile, for (from left to right panels) the fiducial live halo model, the static halo $t=0$ model, and the static halo $t=59t_{\rm d}$ model. 
    Redder profiles indicate earlier times, and bluer profiles indicate later times. 
    The maximum bar strength, $A_{2}^{\rm max}$, is derived from the maximum value of these profiles. 
    The horizontal dashed grey line shows where $A_{2}^{\rm max} > 0.2$, above which we classify the disc as barred. }
    \label{fig:A2radialprofiles}
\end{figure*}

We also identify a complex non-linear relationship with bar formation at different particle resolutions and softening lengths, plotted in Fig.~\ref{fig:softening_scaling}.  
At total resolutions above $N_{\rm d} \geq 5\times10^4$ (with $\mu = 5$) runs with higher softening develop longer, more slowly rotating, weaker bars than runs with smaller softening lengths, until convergence at $N_{\rm d} \geq 10^6$ ($\mu = 5$). 
However, at total resolutions below this threshold, runs with higher softening develop \textit{stronger, longer} bars, often encompassing a significant fraction of the disc mass, while runs with smaller softening lengths experience spurious bar dampening due to collisions between particles.
This phenomenon is easily explained when we consider the importance of disc-halo angular momentum transfer in bar formation: at low resolution and low softening, numerical heating will shift disc particles away from resonant orbits, limiting the exchange of angular momentum, hindering bar growth. 
We do not find convergence with softening length, but suggest that $\epsilon/R_{\rm d} \leq 1/10$ (equivalently $\epsilon \leq z_{\rm d}$) is appropriate. 

\citet{BL2018} showed that simulations with $\epsilon \geq \epsilon_{\rm crit} \equiv \lambda_{\rm crit}/6$ will be artificially stabilised to radial instabilities, where $\lambda_{\rm crit} = 4\pi^2 G\Sigma_{\rm d}(R) / \kappa(R)^2$. 
This indicates that, for a Milky Way-like galaxy, bar formation converges when $\epsilon \leq \epsilon_{\rm crit}$ at $R > 0.1R_{\rm d}$. 
Softening lengths of $\epsilon/R_{\rm d} \leq 1/20$ are required to achieve this. 
Ultimately, \citet{Romeo1994} showed that softening weakens density fluctuations by a factor of $\sim e^{-|\lambda| \epsilon}$, equivalently weakening the contribution to self-gravity, and impacting the bar formation timescale. 
Higher softening increases $t_{\rm bar}$ in a manner similar to increasing $N_{\rm d}$. 

This suggests that if the softening length is too large compared to the scale height of the galaxy, the rate of growth of the initial density fluctuations into the bar mode will be diminished if $N_{\rm d} \geq 5\times10^4$, but enhanced if above this threshold. 
Softening lengths of roughly $\epsilon/R_{\rm d} = 1/20$ produce realistic bar properties. 
Cosmological simulations, which typically set $\epsilon$ to a fixed physical scale, may therefore experience \textit{enhanced} bar formation in low mass galaxies, where bar formation should be limited - if they manage to form bars in such low resolution environments at all. 
Caution must therefore be used when interpreting scaling relations between mass proxies and bar properties in cosmological simulations.

\subsection{The impact of total resolution}
The leftmost column of Fig. \ref{fig:Nresolution} investigates the importance of total resolution with fixed $\mu = 3$. 
Our lowest resolution models (dark red) host discs with $N_{\rm d} = 10^{6}/4$, a number of particles comparable to high mass galaxies in current state-of-the-art cosmological simulations \cite[e.g. a $M_{\rm d} \approx4.6\times10^{11}M_{\odot}$ in \textsc{EAGLE}, or $M_{\rm d}\approx2.1\times10^{10}M_{\odot}$ in \textsc{TNG50}; ][]{Schaye2015,Pillepich2019}. 
The highest resolution models (black line) use $4$ times more particles, with $N_{\rm d} = 10^6$ - such well resolved galaxies are rarely found in cosmological simulations. 
We find that simulations with $N_{\rm d} < 5\times 10^5$ suffer spurious bar dampening after a early bar formation. 
Below this critical threshold, the bar strength declines rapidly in a few $t_{\rm d}$, and the bars that briefly form are too long and rotate too slowly. 
At least $N_{\rm d} \geq 5\times10^5$ disc particles are needed to robustly resolve bar-like structures, in rough agreement with previous convergence studies \citep{Sellwood2003,Dubinski2009,Fujii2018}. 
This implies $N_{\rm h} = 5\times 10^6$ to $10^8$ halo particles are required in a live halo to properly facilitate the angular momentum and energy exchange required for bar formation, also suggested in \citet{Dubinski2009}. 
In general, we find that lower resolution simulations develop weaker, longer, slowly rotating bars after an earlier $t_{\rm bar}$. 
\citet{Ludlow2021} and \citet{Wilkinson2023} have shown that low halo resolution impacts the general kinematics and thickness of disc galaxies due to collisional heating, the rate of which is determined by the individual halo particle mass at fixed halo mass, as in these tests. 
They suggest that at least $N_{\rm h} \geq 10^6$ is required to alleviate these effects, but it is clear now that that the properties of the bar are not actually resolved in such models: bars must be studied in simulations with $N_{\rm h} \geq 5\times 10^6$ per galaxy. 
This implies that far higher resolution than currently implemented in cosmological simulations is required if we wish to properly resolve the bar formation \citep[see, e.g.,][]{Crain2015,Pillepich2019}. 

We calculate the bar formation timescale, $t_{\rm bar}$, for our models with varying $N_{\rm d}$. 
We find that $t_{\rm bar}$ occurs earlier in lower resolution simulations, plotted by arrows in Fig.~\ref{fig:Nresolution}. 
This is expected from the assumption that the bar grows exponentially from minor density fluctuations in the ICs, where the amplitude of these density fluctuations scales like $1/\sqrt{N_{\rm d}}$. 
The amplitude of the density fluctuations thus grows on a timescale equivalent to $\Delta \sim \exp{(t/\tau_{\rm bar})}$ where $\tau_{\rm bar}$ is the exponential bar formation timescale. 
Following the analysis of \citet{Dubinski2009}, we expect the time delay between two simulations with different resolutions, $N_{\rm 0}$ and $N_{\rm 1}$, to be $\Delta t \approx \tau_{\rm bar} \ln{(N_{\rm 1}/N_{\rm 0})}^{1/2}$. 
We fit the live halo data in Fig.~\ref{fig:Nresolution} and find $\tau_{\rm bar} = 0.35t_{\rm bar}$. 
Using the $N_{\rm d}=10^6$ run as a reference point, we calculate the expected delay from the $N_{\rm d} = 10^{6}/4$ run to be $\Delta t = 0.47t_{\rm bar}$, which is close to the measured delay of $0.48t_{\rm bar}$. 
This suggests that the differences in $t_{\rm bar}$ between simulations with different resolution are due primarily to the amplitudes of the initial density fluctuations in the ICs \citep{Dubinski2009}. 

Ultimately, Fig.~\ref{fig:Nresolution} indicates that while numerical effects from $\epsilon/R_{\rm d}$ impact the bar formation process (particularly at poor resolution), these are less important than correctly choosing $N_{\rm d}$ and $N_{\rm h}$, within reasonable limits. 
We find that with $\epsilon/R_{\rm d} \leq 1/10$ and $\mu \leq 10$ simulations of bars are converged, with resolutions $N_{\rm d} \gtrsim 5\times 10^5$ and $N_{\rm h} \gtrsim 5\times10^6$. 

\section{\texorpdfstring{$A_{2}$}{TEXT} Profile Evolution}\label{apx:profiles}
In this work, we use the maximum value of the $m=2$ Fourier amplitude, $A_{2}^{\rm max}$, as a proxy for the strength of the bar. 
For completeness, we present in Fig.~\ref{fig:A2radialprofiles} the full radial $A_{2}$ profiles from which we derive $A_{2}^{\rm max}$ for the fiducial live halo model, the static halo $t=0t_{\rm d}$ model, and the $t=59t_{\rm d}$ static halo model. 
There is a clear (but non-monotonic, given the bar buckling phase) increase in $A_{2}$ over time in the live halo, reflecting the general rise in $A_{2}^{\rm max}$ presented in Fig.~\ref{fig:freeze_halo}. 
No evolution is found in the static halo $t=0t_{\rm d}$ model, consistent with no bar formation. 
Likewise, the $t=59t_{\rm d}$ static halo model shows a general decrease in $A_{2}$ at all $R$ as a function of time, consistent with bar dissolution. 
Importantly, we note the large peak prominence of $A_{2}^{\rm max}$ at most snapshots: this suggests that the bar is the only coherent non-axisymmetric structure forming in our models, and gives us confidence in our measurements of the various bar properties measured from the $A_{2}$ profiles. 

\bsp	
\label{lastpage}
\end{document}